\documentclass[12pt]{iopart}
\usepackage{iopams}
\expandafter\let\csname equation*\endcsname\relax
\expandafter\let\csname endequation*\endcsname\relax
\usepackage{amsmath}
\usepackage{caption}
\usepackage[utf8]{inputenc}
\usepackage[T1]{fontenc}
\usepackage{algpseudocode}
\usepackage{amsfonts}
\usepackage{amssymb}
\usepackage{graphicx}
\usepackage{multirow}
\usepackage{ragged2e}
\usepackage{subfig}
\usepackage{wrapfig}
\usepackage{graphics}
\usepackage{algorithm}
\usepackage[hidelinks]{hyperref}
\usepackage{geometry}
\usepackage{tikz}
\usepackage{mathtools}
\let\svthefootnote\thefootnote
\newcommand\freefootnote[1]{%
  \let\thefootnote\relax%
  \footnotetext{#1}%
  \let\thefootnote\svthefootnote%
}

\providecommand{\keywords}[1]
{
  \small	
  \textbf{\textit{Keywords---}} #1
}

\begin{document}

\title[Encoder Circuit for Surface Code Using MBQC]{Encoder Circuit For Surface Code using
Measurement-Based Quantum Computing Model}

\author{Priyam Srivastava$^{1^{\dagger}}$, Vaibhav Katyal$^{2^{\dagger}}$, 
Ankur Raina$^3$.}
\freefootnote{both authors contributed equally}

\address{$^{1}$ Department of Physics, Indian Institute of Science Education and Research, Bhopal, India.}
\address{$^{2}$ Institute of Theoretical Physics, 
University of G{\"o}ttingen, 37077 G{\"o}ttingen, Germany.}
\address{$^3$ Department of Electrical Engineering and Computer Science, Indian Institute of Science Education and Research, Bhopal, India.}
\ead{priyamsri1498@gmail.com, vaibhavkatyal18@yahoo.com, ankur@iiserb.ac.in}

\begin{abstract}
Surface codes are one of the most important topological stabilizer codes in the theory of quantum error correction. In this paper, we provide an efficient way to obtain surface codes through Measurement-based quantum computation (MBQC) using cluster state as the resource state. Simple two-dimensional surface codes are studied and analyzed using stabilizer formalism. We also present an algorithm to computationally obtain the stabilizer of the surface codes, through which we later determine the distance of the codes. We note the difference in the stabilizers of the surface codes obtained by Fowler \emph{et al.} \cite{Fowler_2012} wherein they used CNOT entangling operation to create the resource state as opposed to the cluster state which is formed using CZ entangling operation. We provide a theoretical calculation to understand this difference. The obtained surface codes can be used practically as an encoder circuit to encode one logical qubit.
\end{abstract}

\maketitle

\section{Introduction}
Quantum information theory \cite{nielsen2001quantum} is a rapidly growing field with the potential to revolutionize computing and cryptography. It offers the prospect of speeding up computational tasks like factorization \cite{shor1999polynomial} and database searching \cite{grover1996fast}, which may be beyond the capabilities of present-day classical computers. However, building a large-scale, reliable quantum computer that can perform practical computations remains a significant challenge. This challenge arises from the complex interactions and correlations between quantum systems, which form the fundamental building blocks of such devices, and their surrounding environment.

These quantum devices rely on entanglement as a core mechanism for interactions. However, the entanglement that mediates these interactions is susceptible to various environmental influences, leading to unwanted errors during computation. Unlike classical information, quantum information is highly fragile and easily corrupted by noise and other sources of interference. For instance, in classical communication, we send a stream of bits (0s and 1s) through a classical channel, where some bits may flip due to noise (i.e., from 0 to 1 or from 1 to 0). Similarly, in quantum communication \cite{Gisin2007Mar}, we send qubits (represented as |0⟩ and |1⟩) through a quantum channel, which also preserves the quantum coherence of the system. In this case, errors can manifest as bit flips, phase changes ($\alpha$, the relative phase between the qubits), or a combination of both. For example, we may encounter the following errors:

\begin{equation}\label{phase+bit} \begin{aligned} 
|0\rangle &\rightarrow |1\rangle,\\
|1\rangle &\rightarrow e^{i\alpha}|1\rangle. \end{aligned} 
\end{equation}

To protect quantum information against such errors, the field of quantum error correction \cite{roffe2019quantum} plays a critical role. Quantum error correction involves developing codes capable of detecting and correcting errors in devices of interest by encoding the information-carrying quantum states into a larger quantum system and using measurements to detect and correct errors. The surface code \cite{Fowler_2012} is one such family of quantum error-correcting codes that can protect quantum information against a wide range of errors. It is based on the principle of stabilizer codes \cite{qecc_ft}. The surface code consists of a two-dimensional lattice of qubits, where each qubit is connected to its four nearest neighbors (assuming a square lattice). The code works by measuring the parity of sets of qubits, known as stabilizers, which form closed loops on the lattice. By repeatedly measuring these stabilizers, errors can be detected and corrected without destroying the quantum information. The encoder circuit for the surface code typically involves connecting or entangling qubits using CNOT gates, which aligns with a gate-based model of computation. In this entangling scheme, control and target qubits are defined based on the measurement pattern followed by a set of qubits called measure qubits in surface codes. The code can correct any single-qubit error and any two-qubit error involving two adjacent lattice qubits. It is desirable because it is relatively simple to implement and can correct multiple errors simultaneously.

Our work proposes a different encoder scheme for implementing such a circuit by employing the resource state used in the measurement-based quantum computation model. In this model, the traditional quantum circuit is replaced by a quantum state called a cluster state, which is prepared by entangling multiple qubits through Ising-type interactions \cite{Briegel2009Jan}. Computation is performed by measuring individual qubits in a specific order and using the measurement outcomes to guide the computation. This model enables fault-tolerant quantum computing schemes \cite{Raussendorf2005Oct} with significantly high error thresholds \cite{PhysRevLett.98.190504}.

In this paper, we aim to study the circuit required to implement surface code computation using a cluster-based MBQC scheme. We begin by introducing different definitions involved in the field of quantum error correction, along with the stabilizer formalism, in Section \ref{qec1}. This foundation will enable us to explore quantum error-correcting principles effectively. In Section \ref{mbqc2}, we delve into the theoretical background of measurement-based quantum computation, providing a comprehensive understanding of how measuring a qubit in the cluster state can control the necessary operations in this computational model. Moving forward, Section \ref{surface3} provides an overview of existing surface code implementation schemes that utilize CNOT entangling operations. It briefly describes the operations involved in implementing such a circuit. Building upon the various concepts discussed in the previous section, we introduce an encoder circuit for implementing similar surface code operations using the measurement-based model in Section \ref{result}. Implementing the surface code using a different scheme involves encoding the quantum information into a large cluster state, followed by an ordered measurement to detect and correct errors. To achieve this, we present an algorithm to realize the stabilizer evolution of our encoder circuit for the surface code and define the stabilizer generators of such a circuit. We implement the algorithm and study the stabilizer evolution of our 3x3 and 5x5 cluster states as examples. We demonstrate the contrast in stabilizer evolution between our 3x3 cluster state with CZ operation and one entangled by CNOT gates. Furthermore, we provide a theoretical study to explain the reason for such a contrast. Finally, we explore the stabilizer evolution of a 5x5 cluster state and define the logical operators and distance of such a circuit to further investigate its error-correcting properties. 
\pagebreak
\\
{\large\textit{Notation}}
\begin{table}[ht!]
    \centering
    \begin{tabular}{c|c}
         \textbf{X}, \textbf{Y}, \textbf{Z} & Measurement in $\sigma_{x}$, $\sigma_{y}$, and $\sigma_{z}$ respectively  \\
         \textbf{X}$_{L}$ & Logical \textbf{X}-Operator \\
         \textbf{Z}$_{L}$ & Logical \textbf{Z}-Operator \\
         $\mathbf{H}$ & Hadamard Gate \\
         $\mathcal{E}$ & Set of all errors \\
         CZ & diag(1,1,1,-1) \\ 
         CZ$_{ij}$ & CZ operation between $i^{\mathrm{th}}$ and $j^{\mathrm{th}}$ qubit\\
         $\mathbb{P}_n$ & Pauli Groups acting on $n$ qubits \\
         $\mathbb{Z}_2$ & Cyclic Group of Order 2 containing elements \{0,1\} \\
         $\mathbb{S}$ & Stabilizer Group \\
         $\mathbb{N}_{\mathbb{S}}$ & Normalizer Group of $\mathbb{S}$ \\
         $\mathcal{H_{\mathbb{S}}}$ & Eigenspace corresponding to $\mathbb{S}$ \\
         $\mathbb{N_{S}\backslash S}$ & Elements in $\mathbb{N_{S}}$ but outside of $\mathbb{S}$\\
         $\mathbb{N_{S}/S}$ & Quotient Group \\
         $\mathcal{M}$ & Measurement Array \\
         $\mathcal{C}$ & Cluster state \\
         $\mathbf{K}^{(j)}$ & Stabilizer of $j^{\mathrm{th}}$ node in a cluster state

    \end{tabular}    
\end{table}
\section{Quantum Error Correction}\label{qec1}
Quantum error correction (QEC) is a procedure to protect the quantum information from unwanted environmental noise \cite{gaitan2008quantum}\cite{suter2016colloquium}. QEC forms one of the most important aspects of quantum computation, and we can safely say that quantum computers would be of no practical use without QEC. Among all the unwanted noise encountered in quantum computation the most prevalent and difficult to correct is decoherence \cite{Schlosshauer2007}. In decoherence, due to the interaction between the environment and the quantum device, quantum information gets nonlocally correlated with the environment. Practically, that means loss of quantum information. Errors due to decoherence degrade the quantum information quickly and render a quantum computer useless. Apart from decoherence, qubit flip and phase error also occurs during quantum computation. Qubit flip error corresponds to applying an \textbf{X}-quantum gate on the state written in the computational basis states $|0\rangle$ and $|1\rangle$. It is analogous to bit flip error in classical error correction. Qubit phase error introduces a relative phase of $\pi$ between the basis states of the qubit and corresponds to applying a \textbf{Z}-quantum gate on the state. From quantum mechanics, we know that, unlike the global phase, relative phases are important as they change the state, and hence the probability density, of the system. There is no analog of phase errors in classical error correction.
\subsection{Difficulties faced while correcting quantum errors}
The first complication we face while correcting quantum errors is that there is no way to protect the quantum information by making extra copies of it, as done in classical coding schemes. No-cloning theorem prohibits making a copy of the qubit thus making quantum error correction difficult to implement.

The second complication is that quantum computers are susceptible to both bit ($\mathbf{X}$)-flip and phase ($\mathbf{Z}$)-flip errors. We should be able to correct both errors simultaneously. But since there is no classical analog of phase flip errors it is hard to devise a quantum error correcting code that allows us to achieve simultaneous correction of both $\mathbf{X}$- and $\mathbf{Z}$-errors.

The last complication is due to the fact that we cannot measure qubits without losing quantum information. The problem of wavefunction collapse due to measurement renders quantum error correction hard to achieve as we would not be able to detect and correct errors without measuring the qubits. We will see in the later sections how stabilizer formalism allows us to overcome this complication.

\subsection{Quantum error-correcting codes}\label{qecc}

Despite the obstacles, it turns out that quantum error correction is possible. Quantum error-correcting codes (QECC) can be viewed as a mapping from $k$ qubits to $n$ qubits, that is, it maps a Hilbert space of $2^k$ dimensions to a Hilbert space of $2^n$ dimensions. The extra $n-k$ qubits allow us to store the information stored in the $k$ logical qubits in a redundant manner, thereby protecting it from errors. The basic principles required to construct a QECC are: 1) To construct a code it is necessary to have an idea of the errors that we want to correct; 2) The principle of redundant information is used in QECC. If a part of the resource is damaged by errors, the logical qubit can still be accurately recovered from the rest of the redundant qubits.

We already saw in the previous section the types of errors that may occur in quantum computation. One of the four possible things can happen to the qubit: nothing ($\mathbf{I}$), qubit flip ($\mathbf{X}$), phase flip ($\mathbf{Z}$), or both ($\mathbf{Y}=i\mathbf{X}\mathbf{Z}$). In an $n$-qubit Hilbert space, the evolution of qubit can be expanded in terms of the operators $\mathbb{P}_n=\left\{\mathbf{I},\mathbf{X},\mathbf{Y},\mathbf{Z}\right\}^{\otimes n}$. Thus, we can express the action of a unitary operator ($\mathbf{U}$) on n-qubits and their environment as 
\begin{equation}
\mathbf{U}:|\psi\rangle\otimes|0\rangle_E\rightarrow\sum_a\mathbf{E}_a|\psi\rangle\otimes|e_a\rangle_E
\end{equation}
where $\mathbf{E}_a\in \mathbb{P}_n$ are the $n$-qubit Pauli operator, $a$ runs over all $2^{2n}$ possible combinations of $\mathbb{P}_n$ and $|e_a\rangle_E$ are the states of the environment. Some important properties of the Pauli group $\mathbb{P}_n$ are:

\begin{itemize}
    \item Each element of $\mathbb{P}_n$ is unitary.
    \item $\forall \mathbf{M}\in\mathbb{P}_n$; $\mathbf{M}^2=\mathbf{I}$ or $\mathbf{M}^2=-\mathbf{I}$
    \item Any two elements $\mathbf{M},\mathbf{N}\in\mathbb{P}_n$ either commute, i.e., $[\mathbf{M},\mathbf{N}]=0$, or anticommute, i.e., $\left\{\mathbf{M},\mathbf{N}\right\}=0$.
\end{itemize} 
To understand how to construct a QECC we will first define some important quantities:
\\
\\
\textbf{Definition 1}: Each Pauli operator can be assigned a weight $t$ which is an integer with $0\leq t\leq n$; the weight is the number of qubits acted upon by non-trivial Pauli operators, that is $\mathbf{X}$, $\mathbf{Y}$, and $\mathbf{Z}$. For example, let us look at some Pauli operators acting on a 4-qubit Hilbert space
\begin{equation}
\begin{aligned}
\mathbf{X}\otimes\mathbf{I}\otimes\mathbf{I}\otimes\mathbf{I}:\;\text{Weight}\; 1,\\
\mathbf{X}\otimes\mathbf{Z}\otimes\mathbf{Z}\otimes\mathbf{I}:\;\text{Weight}\; 3.
\end{aligned}
\end{equation}
Using this definition we see that if we are able to devise a QECC that can identify a subset $\mathcal{E}\subseteq \mathbb{P}_n$ whose elements have weight $t$, then that QECC can correct $t$ errors.
\\
\\
\textbf{Definition 2}: Choose a subgroup $\mathbb{S}\subseteq\mathbb{P}_n$ which satisfies $[\mathbf{M},\mathbf{N}]=0$ $\forall\;\mathbf{M},\mathbf{N}\in\mathbb{S}$. Such a subgroup $\mathbb{S}$ defines an eigenspace $\mathcal{H}_\mathbb{S}\subseteq\mathcal{H}_{2^n}$. This space is called code subspace. Basis vectors of code subspace are protected from errors. Important properties of code subspace are:

\begin{itemize}
    \item If the subgroup $\mathbb{S}$ has $n-k$ generators, then the space $\mathcal{H}_\mathbb{S}$ has $2^k$ dimensions.
    \item Any operator $\mathbf{M}\in\mathbb{S}$ is a stabilizer of any state $|\psi\rangle\in\mathcal{H}_\mathbb{S}$, that is
    \begin{equation}
    \mathbf{M}|\psi\rangle=|\psi\rangle.
    \end{equation}
\end{itemize}
This is why these QECCs are also called stabilizer codes. We will construct QECC in such a way that it is capable of correcting errors $\mathbf{E}_a\in\mathcal{E}$ which anti-commutes with operators from the group $\mathbb{S}$. Therefore, if $\mathbf{M}\in\mathbb{S}$ and $\mathbf{E}_a\in\mathcal{E}$, then for $|\psi\rangle\in\mathcal{H}_\mathbb{S}$ we will have
\begin{equation}
\begin{aligned}
\mathbf{M}\mathbf{E}_a|\psi\rangle&=-\mathbf{E}_a\mathbf{M}|\psi\rangle\\
&=-\mathbf{E}_a|\psi\rangle.
\end{aligned}
\end{equation}
We see that if the eigenvalue of the operator is -1 then the state is affected by the error $\mathbf{E}_a$. Therefore, when constructing a QECC we must proceed as follows: 1) Determine the error operators that we want to correct; 2) From all the possible Pauli operators, choose those which commute with themselves and at the same time anti-commute with the error operators.

Now, a necessary and sufficient condition that must be satisfied by code subspace to make QECC work \cite{knill1997theory} is,
\begin{equation}\label{klcon}
\langle i|\mathbf{E}_a^{\dagger}\mathbf{E}_b|j\rangle=\delta_{ab}\delta_{ij}.
\end{equation}
Here, $|i\rangle$ and $|j\rangle$ are two codewords that belong to $\mathcal{H}_\mathbb{S}$, and $\mathbf{E}_a$ and $\mathbf{E}_b$ are two different errors that we want to correct. Equation \ref{klcon} is also known as the Knill-Laflamme error correction condition. In simple words, the above condition means that different errors should affect states differently. If two different errors change the state in the same way then there is no way to distinguish which error occurred and hence one cannot correct it too. To get a better understanding see Appendix \ref{3qucode} where we have explained, using a simple example, how quantum error correcting codes are devised. 

\subsection{General Stabilizer Codes}\label{genstabcode}
\textbf{Definition 3}: Let $\mathbb{S}$ be the stabilizer group and $\mathcal{H}_\mathbb{S}$ be the corresponding QECC, then we define \cite{qecc_ft}  the normalizer $\mathbb{N}_\mathbb{S}$ of $\mathbb{S}$ in $\mathbb{P}_n$ as 
\begin{equation}
\mathbb{N}_\mathbb{S}=\left\{\mathbf{N}\in\mathbb{P}_n\;\text{s.t.}\; \mathbf{M}\mathbf{N}=\mathbf{N}\mathbf{M} \;\forall\; \mathbf{M}\in \mathbb{S}\right\}.
\end{equation}
Now, if we have an error $\mathbf{E}\in\mathbb{N}_\mathbb{S}$ then our QECC will not be able to detect this error. This is because the eigenvalue of $\mathbf{E}|\psi\rangle\in\mathcal{H}_\mathbb{S}$ when operated with $\mathbf{M}\in \mathbb{S}$ remains +1.
\begin{align*}
\mathbf{M}\mathbf{E}|\psi\rangle=\mathbf{E}\mathbf{M}|\psi\rangle=\mathbf{E}|\psi\rangle.
\end{align*}
Furthermore, errors in $\mathbb{S}$ are not really errors per se, because they leave the encoded state unchanged. This means that our QECC will detect all errors that are either in $S$ or anticommuting with some element of $\mathbb{S}$, that is $\mathbf{E}\in\mathbb{S}\cup(\mathbb{P}-\mathbb{N}_{S})$. In simple words, we can say that QECC only detects errors that are outside of $\mathbb{N}_{S}\backslash \mathbb{S}$.
\\
\\
\textbf{Definition 4}: The distance $d$ of $\mathcal{H}_\mathbb{S}$ is the weight of the smallest Pauli operator $\mathbf{N}$ in $\mathbb{N}_\mathbb{S}\backslash \mathbb{S}$.

A stabilizer code with distance $d$ will correct $\lfloor\frac{d-1}{2}\rfloor$ errors, or in other words, any QECC that corrects $t$ errors has distance 2$t$+1. We can also think of the distance of QECC in the following way: If we look at a subset of $\lfloor\frac{d-1}{2}\rfloor$ qubits, then no information is revealed about the encoded state, that is, the reduced density matrix tells nothing about the relative phases; hence the superposition persists.
Logical operators belong to the normalizer set $\mathbb{N}(S)$. Fowler \emph{et al.} \cite{Fowler_2012} defined the distance of the surface code as the minimum number of physical qubit bit-flips or phase-flips needed to define a logical $\mathbf{X}_L$ or $\mathbf{Z}_L$ operator. We will look into this more deeply in Section \ref{logop}
\\
\textbf{Definition 5}: To define the error syndrome $f(\mathbf{M})$ for a stabilizer code, let $f_M:\mathbb{P}_n\rightarrow\mathbb{Z}_2$
\begin{equation}
f_M(\mathbf{E})=\begin{cases}
0 & \text{if}\;[\mathbf{M},\mathbf{E}]=0\\
1 & \text{if}\;\{\mathbf{M},\mathbf{E}\}=0
\end{cases}.
\end{equation}

Then $f(\mathbf{E})=(f_{M_1}(\mathbf{E}),\dots,f_{M_{n-k}})$, where $M_1,\dots,M_{n-k}$ are the generators of $\mathbb{S}$, is a ($n-k$)-bit binary number. Note that $f(E)$ can have $2^{n-k}$ different values. Two different errors $\mathbf{E}$ and $\mathbf{F}$ have the same error syndrome if they commute with the same set of generators of $\mathbb{S}$. Then, this would mean that $\mathbf{E}^{\dagger}\mathbf{F}\in\mathbb{N}_\mathbb{S}$. The error syndrome can distinguish different errors if $\mathbf{E}^{\dagger}\mathbf{F}\notin\mathbb{N}_\mathbb{S}$. Further, if $\mathbf{E}^{\dagger}\mathbf{F}\in \mathbb{S}$ then $\mathbf{E}$ and $\mathbf{F}$ act in the same way on the codeword. Hence, there is no need to distinguish them. Therefore, we can conclude that the QECC corrects errors for which $\mathbf{E}^{\dagger}\mathbf{F}\notin\mathbb{N}_\mathbb{S}\backslash \mathbb{S}$ for all pairs $(\mathbf{E},\mathbf{F})$. From this condition, we can see how the distance of QECC is $2t+1$. If our QECC  corrects $t$ errors the product $E^{\dagger}F$ can act on at most 2$t$ qubits. Since to correct the errors $\mathbf{E}^{\dagger}\mathbf{F}$ must be outside $\mathbb{N}_\mathbb{S}\backslash \mathbb{S}$, we will have the weight of smallest Pauli operator in $\mathbb{N}_\mathbb{S}\backslash \mathbb{S}$, which is defined as the distance, as 2$t$+1.  Now, in order to perform an error correction operation, all we need to do is measure the eigenvalue of each generator of the stabilizer. For a non-degenerate code, error syndrome will be different for different errors and hence will exactly tell which error to correct. Unlike a non-degenerate code, for a degenerate code, there exist pairs of errors $(\mathbf{E},\mathbf{F})$ that have the same error syndrome. This occurs if $\mathbb{S}$ has elements of weight less than distance $d$.

We know that elements of $\mathbb{N}_\mathbb{S}$ commute with elements of the $\mathbb{S}$, in fact, $\mathbb{S}$ lies inside $\mathbb{N}_\mathbb{S}$. This means $\mathbb{N}_\mathbb{S}$ moves codewords around within $\mathcal{H}_\mathbb{S}$, and hence we can interpret it as encoded operations on the codewords. Only elements in $\mathbb{N}_\mathbb{S}/ \mathbb{S}$ acts non-trivially on $\mathcal{H}_\mathbb{S}$. Here, $\mathbb{N}_\mathbb{S}/\mathbb{S}$ is the quotient or factor group, which is the set of cosets of $\mathbb{S}$ in $\mathbb{N}_\mathbb{S}$, and is defined as
\begin{equation}
\mathbb{N}_\mathbb{S}/\mathbb{S}=\left\{\mathbb{S}, a_1\mathbb{S}, a_2\mathbb{S}, \dots, a_m\mathbb{S}\right\}.
\end{equation}

Note that we can only define quotient group $\mathbb{N}_\mathbb{S}/\mathbb{S}$ when $\mathbb{S}$ is an invariant subgroup of $\mathbb{N}_\mathbb{S}$. This is precisely the case here as, by definition, $\mathbb{N}_\mathbb{S}$ commutes with every element of $\mathbb{S}$, and hence the right cosets ($\mathbb{S}a_i\;\forall\;a_i\in\mathbb{N}_\mathbb{S}\backslash \mathbb{S}$) will be the same as left cosets ($a_i\mathbb{S}\;\forall\;a_i\in\mathbb{N}_\mathbb{S}\backslash \mathbb{S}$). It is possible to show that this group is a Pauli group with size $k-n-r$.
Putting these considerations together we can define an automorphism  $\mathbb{N}_\mathbb{S}/\mathbb{S}\rightarrow\mathbb{P}_k$. We will write \textbf{X}$_{i}$ and \textbf{Z}$_{i}$, where $i=1,\dots,k$, as the encoded/logical operators. \textbf{X}$_{i}$ maps to \textbf{X} in $\mathbb{P}_k$ and  \textbf{Z}$_{i}$ maps to \textbf{Z} in $\mathbb{P}_k$.

\section{Measurement Based Quantum Computation}\label{mbqc2}

Measurement-based quantum computation (MBQC), which originated from the pioneering work of Raussendorf and Briegel \cite{raussendorf2001one}, is a technique to perform quantum computation using only local measurements. The central physical ingredient of MBQC is the cluster state \cite{briegel2001persistent}. It is a special type of entangled state that allows for quantum computation using only one-qubit projective measurements. This scheme is also called the "one-way quantum computer" (1WQC) because the entanglement in the cluster state is destroyed
by the one-qubit measurements and therefore it can only
be used once. Before we go into the details, let us have a general picture of how 1WQC works.

\begin{figure}[ht!]
\centering
\includegraphics[width=65mm]{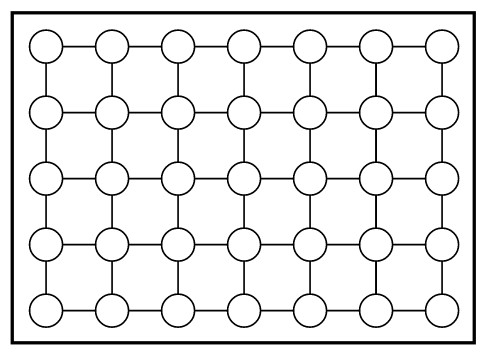}
\captionof{figure}{Two dimensional cluster state. Vertices represent qubits and the edges represent entanglement between the adjacent qubits.}
\label{2d_cs}
\end{figure}

The first step to 1WQC is to form the cluster state. Figure \ref{2d_cs} shows a 2d cluster state. A cluster state can be created efficiently in any system using Ising-type interaction at very low temperatures. Next, we write the information onto the cluster through entanglement. Then, we apply local one-qubit measurements on the cluster state qubits to process the information. The processed information is then read out. We see that cluster state provides in advance all entanglement that is involved in the subsequent
quantum computation. It is important, at this stage, to distinguish between cluster qubits and logical qubits. The logical qubits constitute the quantum information being processed, and cluster qubits are the qubits that form the cluster state. Measurements in the \textbf{X} - (and \textbf{Y}) basis propagate the quantum information through the circuit i.e., they act as a wire, and measurement in \textbf{Z}-basis removes the respective qubit from the cluster (Section \ref{effect_z}). The processing is finished once all qubits except the last one on each wire have been measured. The remaining unmeasured qubits form the
quantum register which is then read out. Results of previous measurements determine in which basis the output qubits need to be measured for the final readout.

\subsection{Cluster States}
The most important ingredient of MBQC is the cluster state. Cluster state is a special type of entangled state that allows for universal quantum computation using local measurement \cite{raussendorf2012quantum}. A 2$D$ cluster state, as shown in Figure \ref{2d_cs}, can be thought of as a blank slate upon which we can draw the quantum circuit using local measurements. Cluster states are a particular case of graph states. For a particular graph $\mathcal{C}$ = ($C$, E), where $C$ are the vertices and $E$ edges; we use the following algorithm to generate cluster states: 1) To each vertex $i\in C$ we associate the qubit $|+\rangle$; 2) Next all the pairs of neighboring qubits $(i,j)\in E$ are entangled by the CZ operation which is defined as
\begin{equation}\label{cpdef}
    \text{CZ}_{ij}=|0\rangle_i\langle0|\otimes\mathbf{I}_j+|1\rangle_i\langle1|\otimes\mathbf{Z}_j.
\end{equation}
As a result, we can write the cluster state $\mathcal{C}$ as
\begin{equation}
    |G\rangle_C=\prod_{(i, j) \in E} \text {CZ}_{ij} \bigotimes_{i \in V}|+\rangle_{i}.
\end{equation}
The cluster state $|G\rangle_C$ satisfies the following set of eigenvalue equations
\begin{equation}\label{cs_eigenvalue}
\left(\mathbf{X}_i\prod_{j\in N(i)}\mathbf{Z}_j\right)|G\rangle_C=(-1)^{\lambda}|G\rangle_C,
\end{equation}
where $N(i)$ is the set of neighbouring vertices of vertex $i$.
Also, $\lambda\in \{0,1\}\;\forall\; a\in \mathcal{C}$. We define $K^{(a)}$ 
\begin{equation}\label{stabilizers_cs}
K^{(a)}=\mathbf{X}_a\prod_{j\in N(a)}\mathbf{Z}_j
\end{equation}
as the stabilizers of the cluster state. We notice that stabilizers corresponding to different sites commute with each other. This means there exist a set of common eigenstates. All the eigenstates form a group and every one of them is equally good for computation. Eigenstates within this group with the same set of eigenvalues $\{\kappa\}$ form an equivalence class. There are $2^{|C|}$ such equivalence classes corresponding to every $2^{|C|}$ eigenvalue sets $\{\kappa\}$. We can choose any one of the equivalence classes and select any one element from it as its representative. $2^{|C|}$ such representatives from every equivalence class forms the  |$C$|-qubit Hilbert space. Note that elements of the equivalence class are orthogonal to each other \cite{raussendorf2003measurement}.

There are essentially two ways to create a cluster state. The first way is to measure all the stabilizers $\mathbf{K}^{(i)}$ given in Equation \ref{stabilizers_cs} for $i\in C$ on an arbitrary $|C|$-qubit state or to cool into the ground state of the Hamiltonian $\hat{\mathbf{H}}_K=-\hbar g\sum_{i\in C}\kappa_a\mathbf{K}^{(i)}$. A more practical way is to first create a product state $|+\rangle_C=\bigotimes_{i\in C}|+\rangle_i$. Then we apply a unitary transformation $\mathbf{S}^{(C)}$, which is given as
\begin{equation}\label{unitary_int}
\mathbf{S}^{(C)}=\prod_{a,b\in C|b-a\in\gamma_D}\mathbf{S}^{ab},
\end{equation}
on the product state $|+\rangle_C$. $\mathbf{S}^{ab}$ has the form
\begin{equation}
\mathbf{S}^{ab}=\dfrac{1}{2}\left(\mathbf{I}+\mathbf{Z}^{(a)}+\mathbf{Z}^{(b)}-\mathbf{Z}^{(a)}\otimes\mathbf{Z}^{(b)}\right).
\end{equation}
The condition $b-a\in\gamma_D$ in Equation \ref{unitary_int} restricts the interaction of a qubit with only its nearest neighbors. Therefore, in dimension ($D=1,2,3$), we will have  $\gamma_1=\{1\}$, $\gamma_2=\{(1,0)^T,(0,1)^T\}$, and $\gamma_3=\{(1,0,0)^T,(0,1,0)^T,(0,0,1)^T\}$. 
\subsection{Removing redundant cluster qubits}\label{effect_z}

There always exist qubits on the cluster state that are not needed in realizing a quantum circuit. We can remove these redundant qubits by measuring them in the $\mathbf{Z}-$ basis. The resulting entangled state is again a cluster state. To see this we write the cluster state after measurement as a product of entangled quantum state on the cluster $C_N$ of the unmeasured qubits and product state on cluster $C\backslash C_N$ of qubits which are measured. Therefore,
\begin{equation}
|G\rangle_C\longrightarrow |Z\rangle_{C\backslash C_N}\otimes |G^{\prime}\rangle_{C_N},
\end{equation}
where $|Z\rangle_{C\backslash C_N}=\bigotimes_{i\in C\backslash C_N}|s_i\rangle_{i,z}$ and $s_i$ are the results of the $\sigma_z-$measurements. $|G\rangle_C$ is the initial cluster state. We notice that we can also write $|Z\rangle_{C\backslash C_N}\otimes |G^{\prime}\rangle_{C_N}$ as follows

\begin{equation}
|Z\rangle_{C\backslash C_N}\otimes |G^{\prime}\rangle_{C_N}=\left(\bigotimes_{i\in C\backslash C_N} \dfrac{1+(-1)^{s_i}\mathbf{Z}^{(i)}}{2}\right)|G\rangle_C.
\end{equation}

The term in the above equation is the projection operator $\mathcal{P}$ corresponding to the $\mathbf{Z}-$measurements. Since $|G\rangle_C$ satisfies Equation \ref{cs_eigenvalue} we insert $K^{(i)}$ with $i\in C_N$ in the r.h.s of the above equation between the projector and the state, and simplify it to obtain
\begin{equation}\label{remove_z}
\begin{aligned}
|Z\rangle_{C\backslash C_N}\otimes |G^{\prime}\rangle_{C_N}&=(-1)^{\lambda}\left(\bigotimes_{i\in C\backslash C_N} \dfrac{1+(-1)^{s_i}\mathbf{Z}^{(i)}}{2}\right)K^{(i)}|G\rangle_C\\
&=(-1)^{\lambda}\mathbf{X}^{(i)}\bigotimes_{b\in N(i)\cap C_N}\mathbf{Z}^{(b)}\bigotimes_{b\in N(i)\cap C\backslash C_N}\mathbf{Z}^{(b)}|Z\rangle_{C\backslash C_N}\otimes |G^{\prime}\rangle_{C_N}\\
&=(-1)^{\lambda^{\prime}}K^{\prime (a)}|Z\rangle_{C\backslash C_N}\otimes |G^{\prime}\rangle_{C_N}.
\end{aligned}
\end{equation}
with the new stabilizers of the remaining unmeasured qubits 
\begin{equation}
K^{\prime (a)}=\mathbf{X}^{(a)}\bigotimes_{b\in N(a)\cap C_N}\mathbf{Z}^{(b)}.
\end{equation}
and the set $\{\kappa_a^{\prime}\}$ specifying the eigenvalues 
\begin{equation}
\lambda^{\prime}=(\lambda+\sum_{b\in N(a)\cap C\backslash C_N}s_b)\;\text{mod}\;2.
\end{equation}
To obtain the second equality in Equation \ref{remove_z} we used the fact that projection operator $\mathcal{P}$ and correlation operator $K^{(a)}$ commute with each other as they act on different qubits. To obtain the last equality we apply $\bigotimes_{b\in N(a)\cap C\backslash C_N}\mathbf{Z}^{(b)}$ on $|Z\rangle_{C\backslash C_N}$ to get $(-1)^{\sum_{b\in N(a)\cap C\backslash C_N}s_b}|Z\rangle_{C\backslash C_N}$. The new correlation operators $K^{\prime (a)}$ act only on the cluster qubits in $C_N$ and obey the following eigenvalue equations as can be seen from Equation \ref{remove_z} 
\begin{equation}
K^{\prime (a)}|G^{\prime}\rangle_{C_N}=(-1)^{\lambda^{\prime}}|G^{\prime}\rangle_{C_N},\quad\forall\;a\in C_N.
\end{equation}
The above equation proves that the residual state $|G^{\prime}\rangle_{C_N}$ is again a cluster state. We also note that the measurement results of the removed qubits influence the process of computation. Since, any cluster state $|G^{\prime}\rangle_{C_N}$ is equally good for computation we can put $\lambda^{\prime}=0$. The above analysis shows that we can discard the redundant qubits by measuring them in the $\mathbf{Z}-$ basis.

\subsection{Scheme to realize a gate on a cluster state}\label{schemembqc}

To implement a gate $g$ we take a cluster $\mathcal{C}(g)$. We divide it into three sections; an input section $\mathcal{C}_I(g)$, a body $\mathcal{C}_B(g)$ and an output section $\mathcal{C}_O(g)$ such that 

\begin{equation}
\begin{aligned}
\mathcal{C}_{I}(g) \cup \mathcal{C}_{B}(g) \cup \mathcal{C}_{O}(g)&=\mathcal{C}(g), \\
\mathcal{C}_{I}(g) \cap \mathcal{C}_{B}(g) &=\emptyset, \\
\mathcal{C}_{I}(g) \cap \mathcal{C}_{O}(g) &=\emptyset, \\
\mathcal{C}_{B}(g) \cap \mathcal{C}_{O}(g) &=\emptyset.
\end{aligned}
\end{equation}
The input qubits in $\mathcal{C}_{I}(g)$ are prepared in the state $\left|\psi_{\text {in }}\right\rangle$ on which we need to apply the gate $g$, while the qubits in $\mathcal{C}_B(g) \cup \mathcal{C}_O(g)$ are in the usual $|+\rangle=|0\rangle_x$ state. We entangle the qubits via the interaction $\mathbf{S}^{(\mathcal{C}(g))}$ given in equation \ref{unitary_int}. To realize the gate on the state $ \left|\psi_{\text {in }}\right\rangle$ appropriate measurements are applied to the qubits in $\mathcal{C}_B(g)$. It was seen that apart from required operation $\mathbf{U}_g$ corresponding to the gate $g$ obtained on the output qubits $\mathcal{C}_O(g)$, we get an extra Pauli operation $\mathbf{U}_{e}$ which depends on the measurement outcomes of the qubits measured in $\mathcal{C}_B(g)$ \cite{raussendorf2003measurement}, that is, 
\begin{equation}
\left|\psi_{\text {out }}\right\rangle=\mathbf{U}_{e} \mathbf{U}_g\left|\psi_{\text {in }}\right\rangle ,
\end{equation}
where
\begin{equation}
\mathbf{U}_{e}=\bigotimes_{i=1}^n\left(\mathbf{X}^{[i]}\right)^{x_i}\left(\mathbf{Z}^{[i]}\right)^{z_i}.
\end{equation}
Here, $n=|I|=|O|$ and $\{\mathbf{X},\mathbf{Z}\}^{[i]}$ denotes Pauli operator acting on the logical qubit $i$ and not the cluster qubits. The values $x_i,\; z_i \in \{0,1\}$ are computed from the measurement outcomes $\{s_k|k\in \mathcal{C}_I(g)\cup\mathcal{C}_B(g)\}$.

Now, we define a theorem that provides a criterion describing the functioning of the gate in MBQC.
\\
\\
\textbf{Theorem 1}: Let $\mathcal{C}(g)=\mathcal{C}_{I}(g) \cup \mathcal{C}_{B}(g) \cup \mathcal{C}_{O}(g)$ be a cluster used to implement a unitary transformation $\mathbf{U}$ corresponding to gate $g$, and $|\phi\rangle_{\mathcal{C}_g}$ be the cluster state. Suppose the state $|\psi\rangle_{\mathcal{C}_g}=P_{\{s\}}^{\left(\mathcal{C}_B(g)\right)}(\mathcal{M})|\phi\rangle_{\mathcal{C}(g)}$ obeys the 2$n$ eigenvalue equations for the measurement pattern $\mathcal{M}$,

\begin{equation}
\begin{aligned}
& \mathbf{X}^{\left(\mathcal{C}_I(g), i\right)}\left(\mathbf{U} \mathbf{X}^{(i)} \mathbf{U}^{\dagger}\right)^{\left(\mathcal{C}_O(g)\right)}|\psi\rangle_{\mathcal{C}(g)}=(-1)^{\lambda_{x, i}}|\psi\rangle_{\mathcal{C}(g)} ,\\
& \mathbf{Z}^{\left(\mathcal{C}_{I(g), i)}\right.}\left(\mathbf{U} \mathbf{Z}^{(i)} \mathbf{U}^{\dagger}\right)^{\left(\mathcal{C}_O(g)\right)}|\psi\rangle_{\mathcal{C}(g)}=(-1)^{\lambda_{z, i}}|\psi\rangle_{\mathcal{C}(g)},
\end{aligned}
\end{equation}
with $\lambda_{x,i},\lambda_{z,i}\in \{0,1\}$ depends on the measurement outcomes of qubits in $\mathcal{C}_{M}(g)$ which are measured according to $\mathcal{M}$ and $1\leq i \leq n$. $n$ is the number of input qubits. We can, then, realize gate $g$ on the cluster state using Scheme \ref{schemembqc} with measurement pattern described by $\mathcal{M}$, and the measurements of the qubits in $\mathcal{C}_I(g)$ being \textbf{X} measurement. After measurement, we can relate input and output states as
\begin{equation}
    |\psi_{out}\rangle=\mathbf{U}\mathbf{U}_{\Sigma}|\psi_{in}\rangle,
\end{equation}
where $\mathbf{U}_{\Sigma}$ is the byproduct operator given by
\begin{equation}
    \mathbf{U}_{\Sigma}= \bigotimes_{\left(\mathcal{C}_I(g) \ni i\right)=1}^n\left(\mathbf{Z}^{[i]}\right)^{s_i+\lambda_{x, i}}\left(\mathbf{X}^{[i]}\right)^{\lambda_{z, i}},
\end{equation}
where $s_i\in\{0,1\}$ is the measurement outcome of $\mathbf{X}$ measurement performed on the $\mathcal{C}_{I}(g)$.

In the next section, we have implemented the theory we have developed till now to implement surface codes on $3\times3$ and $5\times5$ cluster states. Before we start discussing our work we would like to state an important lemma that would be helpful during our discussion on stabilizer evolution in the later sections
\\
\\
\textbf{Lemma 1} If $\mathbb{S} = \left\{\mathbf{S}_i\right\}_{i=1}^m$ are the stabilizers of a state $|\psi\rangle$ then $\left\{\mathbf{U}\mathbf{S}_i\mathbf{U}^{\dagger}\right\}_{i=1}^m$ are the stabilizers of the state $\mathbf{U}|\psi\rangle$, where $\mathbf{U}$ is an unitary operator.

\section{Surface Code implementation}\label{surface3}
A surface code is a special type of stabilizer code. The idea behind surface codes is to encode information in "homological degrees of freedom" (See \cite{bombin2013introduction} and \cite{dan_b} to understand the topology of surface codes). See Appendix \ref{homology} for a brief introduction to homology. The surface code implementation, as done by Fowler \emph{et al.} \cite{Fowler_2012}, utilizes an array of states initialized in the ground state (computational basis) followed by a patterned CNOT implementation based on whether a $\mathbf{Z}$-measurement or $\mathbf{X}$-measurement takes place on the measure qubit. However, the work demonstrated in this paper follows a slightly different pattern as the qubits are initialized in the  |+$\rangle$ state instead of the computational basis. The cluster of qubits initialized in such a manner is then entangled using the symmetric CZ gate instead of the previously mentioned CNOT which is not symmetric. 

\begin{figure}[ht!]
\centering
\includegraphics[width=\textwidth]{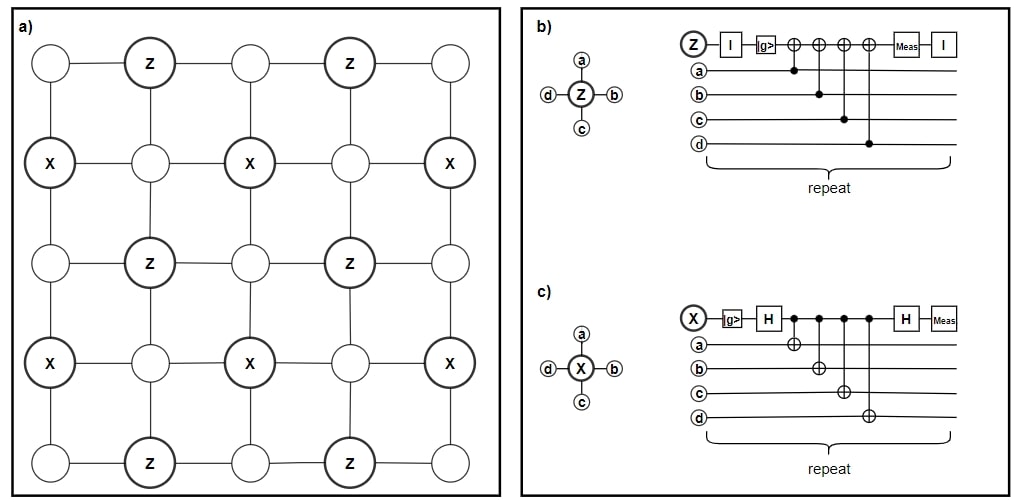}
\captionsetup{width=0.9\linewidth}
\caption{(a) A two-dimensional array implementation of surface code. (b) Quantum circuit for one surface code
cycle for a measure-X qubit, which stabilizes $\mathbf{X}_a\mathbf{X}_b\mathbf{X}_c\mathbf{X}_d$. (c) Quantum circuit for one surface code cycle for a measure-Z qubit, which stabilizes $\mathbf{Z}_a\mathbf{Z}_b\mathbf{Z}_c\mathbf{Z}_d$.\cite{Fowler_2012}}
\label{2d_surface_code}
\end{figure}

But before we discuss our work we will briefly talk about Fowler's way of implementing surface codes. Figure \ref{2d_surface_code} shows the surface code on a 2D array. Open circles represent data qubits and filled circles are the measurement qubits. The measurement qubits, essentially, are used to stabilize and manipulate the quantum state of the data qubits. There are two types of measurement qubits: `measure-Z' qubits (colored green), and `measure-X' qubits (colored orange). Each data qubit is coupled to two measure-Z qubits and two measure-X qubits, and each measurement qubit is coupled with four data qubits.

All the measurement qubits are initialized in ground state $|0\rangle$. Figure \ref{2d_surface_code}b and c show one complete cycle for measure-X and measure-Z qubits. After projective measurement of all measure qubits, the state $|\psi\rangle$ of all data qubits simultaneously satisfies $\mathbf{Z}_a\mathbf{Z}_b\mathbf{Z}_c\mathbf{Z}_d|\psi\rangle=Z_{abcd}|\psi\rangle$ with eigenvalues $Z_{abcd}=\pm 1$, and $\mathbf{X}_a\mathbf{X}_b\mathbf{X}_c\mathbf{X}_d|\psi\rangle=X_{abcd}|\psi\rangle$ with eigenvalues $X_{abcd}=\pm 1$. The state $|\psi\rangle$ is called the \emph{quiescent state}. In Figure \ref{2d_surface_code}, there are 12 measure qubits. This means there are $2^{12}$ possible measurement outcomes. Corresponding to each outcome there will be quiescent state $|\psi\rangle$. We randomly select any one state $|\psi\rangle$ from $2^{12}$ possible states after one complete cycle as shown in Figure \ref{2d_surface_code}b and c. Once selected the same state $|\psi\rangle$ will be maintained by each subsequent cycle of the sequence unless an error occurs. This is because all the stabilizers commute with each other. The commutation relation of stabilizers with no common data qubits is trivial to show. Here we show the commutation relation of the stabilizers which have two common data qubits
\begin{equation}
\begin{aligned}
{\left[\mathbf{X}_{a} \mathbf{X}_{b} \mathbf{X}_{c} \mathbf{X}_{d}\right.} \left., \mathbf{Z}_{a} \mathbf{Z}_{b} \mathbf{Z}_{e} \mathbf{Z}_{f}\right] 
&=\left(\mathbf{X}_{a} \mathbf{Z}_{a}\right)\left(\mathbf{X}_{b} \mathbf{Z}_{b}\right) \mathbf{X}_{c} \mathbf{X}_{d} \mathbf{Z}_{e} \mathbf{Z}_{f} \\
&=\left(\mathbf{Z}_{a} \mathbf{X}_{a}\right)\left(\mathbf{Z}_{b} \mathbf{X}_{b}\right) \mathbf{X}_{c} \mathbf{X}_{d} \mathbf{Z}_{e} \mathbf{Z}_{f} \\
&=0.
\end{aligned}
\end{equation}

We can visualize the 2D surface code given in Figure \ref{2d_surface_code} as a set of data qubits on every edge of an embedded square lattice on the planar surface, and measure qubits on every vertex (face) and face (vertex) of the lattice (dual lattice).
\subsection{Single qubit error detection and correction}
Let us suppose that a bit-flip error occurs on one of the data qubits. The two adjacent measure-Z qubits will detect this because the sign of these measure-Z qubits will change. This will also change the quiescent state
\begin{align}
\mathbf{Z}_{a} \mathbf{Z}_{b} \mathbf{Z}_{c} \mathbf{Z}_{d}(\mathbf{X}_a|\psi\rangle)&=-\mathbf{X}_a(\mathbf{Z}_{a} \mathbf{Z}_{b} \mathbf{Z}_{c} \mathbf{Z}_{d}|\psi\rangle)\\
&=-Z_{abcd}(\mathbf{X}_a|\psi\rangle).
\end{align}

This shows that $\mathbf{X}_a|\psi\rangle$ is an eigenstate of $\mathbf{Z}$ stabilizer but with an opposite sign. This will be more clear with the following example:
Let us assume that a bit-flip error occurs on the $7^{\text{th}}$ data qubit. We select the quiescent state $|\psi\rangle$ corresponding to the set $\{1,-1,-1,1,1,\cdots\}$, which has measurement outcomes of all the 12 measure qubits. Since, bit-flip error occurs on the $7^{\text{th}}$ data qubit, sign of the measurement outcome at $6^{\text{th}}$ and $8^{\text{th}}$ measure-Z qubits will change. Similarly, we can detect phase-flip errors. If instead of a bit-flip error on 7$^{\text{th}}$ data qubit a phase-flip error occurs then measurement outcome at  $4^{\text{th}}$ and $9^{\text{th}}$ measure-X qubit will flip signs.

Now, to correct for the bit-flip we apply a second $\mathbf{X}_a$ operator on the corrupted data qubit. The same can be done for the phase-flip errors where we apply $\mathbf{Z}_a$ on the data qubit to correct for the error. An issue with applying an extra gate operation for practical purposes is that we cannot obtain $100\%$ fidelity. This could introduce more errors in the surface code. Instead, the errors are corrected using a classical control software which changes the sign of measurement qubits that get flipped due to errors.

\subsection{Logical Operators}\label{logop}

\begin{figure}[ht!]
\centering
\includegraphics[width=75mm]{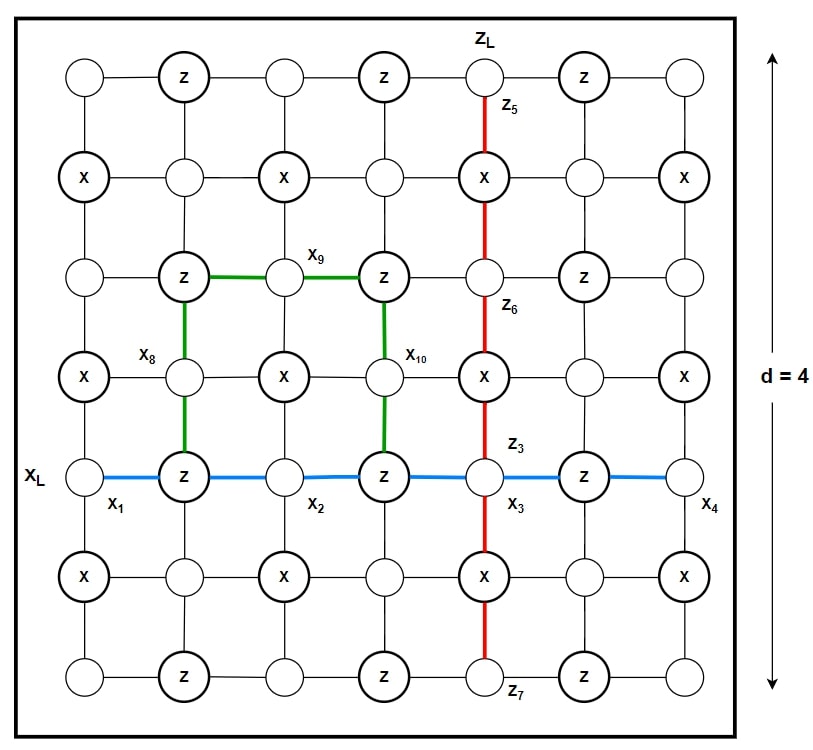}
\captionsetup{width=0.9\linewidth}
\caption{A square 2D array of data qubits, with
X boundaries on the left and right, and Z boundaries on the top and bottom. The array has 25 data qubits, but only 24 $\mathbf{X}$ and $\mathbf{Z}$ stabilizers. A product chain $\mathbf{X}_L=\mathbf{X}_1\mathbf{X}_2\mathbf{X}_3\mathbf{X}_4$ of $\mathbf{X}$ operators connects the two X boundaries, commutes with all the array stabilizers and changes the array state from the quiescent state $|\psi\rangle$ to $|\psi_X\rangle=\mathbf{X}_L|\psi\rangle$ with the same measurement outcomes as $|\psi\rangle$. A second product chain $\mathbf{Z}_L=\mathbf{Z}_5\mathbf{Z}_6\mathbf{Z}_3\mathbf{Z}_7$ connects the two Z boundaries and commutes with the array stabilizers; it changes the array state from $|\psi\rangle$ to $|\psi_Z\rangle=\mathbf{Z}_L|\psi\rangle$. The operator chains $\mathbf{X}_L$ and $\mathbf{Z}_L$ anti-commute.}
\label{logical_operator_sc}
\end{figure}

A planar 2D surface code with a single boundary cannot encode any qubit. This is because all the possible 1-chains are homologically equivalent. Therefore, the first homology group $\mathbb{H}_1$ will be isomorphic to the trivial group $\mathbb{Z}_1$. Since it does not have any generators there will be no encoded qubit. It may seem that the surface code in Figure \ref{2d_surface_code} can also not encode any qubit, but that is not the case. This is because it has two different kinds of boundaries. There is a boundary that terminates with measure-X qubits, which in literature is referred to as \emph{smooth boundary}. The other boundary terminates with a measure-Z qubit, called \emph{rough boundary}. Since there are two different types of boundaries the first homology group  $\mathbb{H}_1$ will be isomorphic to $\mathbb{Z}_2$. Since $\mathbb{Z}_2$ has one generator, the surface code will encode 1 qubit. This can also be seen in Figure \ref{logical_operator_sc}. It has 25 data qubits and 24 measure qubits, so there are $2\times 25$ degrees of freedom in the data qubits and $2\times 24$ constraints from the stabilizer measurements. The two unconstrained degrees of freedom act as a logical or encoded qubit \cite{Fowler_2012}\cite{dennis2002topological}.  

We can now define logical operators that allow us to manipulate these additional degrees of freedom. Logical operators form closed 1-chain cycles. In Figure \ref{logical_operator_sc}  operators $\mathbf{X}_L$ (in blue) and $\mathbf{Z}_L$ (in red) show the two possible logical operators acting on the data qubits. We note that the product chain $\mathbf{X}_L$ and $\mathbf{Z}_L$ is closed 1-cycles and homologically inequivalent (See Appendix \ref{homology} to read more on the homology of curves). They represent the two equivalence classes of $\mathbb{H}_1$. We can choose other chains of single qubit operator products to define different $\mathbf{X}_L$ and $\mathbf{Z}_L$. Consider for example the chain $\mathbf{X}_L^{\prime}=\mathbf{X}_1\mathbf{X}_{8}\mathbf{X}_{9}\mathbf{X}_{10}\mathbf{X}_3\mathbf{X}_4$ also shown in Figure \ref{logical_operator_sc}. It can be easily seen that it is homologically equivalent to $\mathbf{X}_L$ and hence belong to the same equivalence class (See the Section \ref{homological_equivalence}). Notice that we can write $\mathbf{X}_L^{\prime}=(\mathbf{X}_2\mathbf{X}_{8}\mathbf{X}_{9}\mathbf{X}_{10})\mathbf{X}_L$. The term in the parentheses is the stabilizer operator. Therefore, if we operate on a quiescent state $|\psi\rangle$ with $\mathbf{X}_L^{\prime}$ we will get the same state obtained when $\mathbf{X}_L$ is applied on $|\psi\rangle$ up to an overall phase. Hence, there is only one linearly independent $\mathbf{X}_L$ operator for this array.

At first glance this result may seem mysterious; after all, if we operate on the array with a loop of $\mathbf{X}$ operators corresponding to a stabilizer $\mathbf{X}_{\text{loop}}=\mathbf{X}_2\mathbf{X}_{8}\mathbf{X}_{9}\mathbf{X}_{10}$ we are bit-flipping four data qubits so the state $|\psi\rangle^{\prime}=\mathbf{X}_{\text{loop}}|\psi\rangle$ should be different from $|\psi\rangle$. The only way $|\psi\rangle$ and $|\psi\rangle^{\prime}$ are the same is if $|\psi\rangle$ already contains the superposition of un-bit-flipped and bit-flipped data qubits. This is precisely how we define surface code. A surface code state is the superposition of all homologically equivalent cycles. A stabilizer loop like $\mathbf{X}_{\text{loop}}$ will only lead to homologically equivalent configurations. Look at Appendix \ref{homology} to get a mathematical perspective and intuitive understanding of the homology of curves.

Notice that the $\mathbf{Z}_L$ operator chain is in the dual lattice, which makes it a 1-cochain. Since in the dual lattice we have measure-X qubits (i.e., $|+\rangle$ and $|-\rangle$ are the basis states); the face stabilizer will be a loop of $\mathbf{Z}$ operators. So, we see that the same argument applies to the $\mathbf{Z}_L$ operator.

\subsubsection{Creating logical qubits:}

To increase the number of logical qubits we need to define more non-trivial homologically inequivalent cycles. An efficient way to do that is to create holes inside the boundaries of the array. This can be done in two ways: turn off measure-Z qubits or measure-X qubits.

When we create a hole inside the array by turning off a measure-Z qubit, we create additional two degrees of freedom or another logical qubit. This logical qubit is called a Z-cut qubit.  The newly formed $\mathbf{X}_L$ logical operator connects the array's outer X-boundary with the internal X-boundary of the hole. The $\mathbf{Z}_L$ logical operator forms a loop (joining measure-X qubits) around the Z-cut hole. Note that this loop is not the stabilizer because it is not simply connected. It is clear that this hole/defect is in the dual basis, we call the single Z-cut qubit a \emph{smooth defect} or a \emph{dual defect}.

Analogously, we can define an X-cut qubit by turning off the measure-X qubit. Then $\mathbf{Z}_L$ operator is a chain of $\mathbf{Z}$ operators from the array Z boundary to the internal Z boundary created by turning off the measure-X qubit, and $\mathbf{X}_L$ operator is a loop of $\mathbf{X}$ bit-flips surrounding the X-cut hole. A single X-cut qubit is called a \emph{rough defect} or a \emph{primal defect}.
We can create more logical qubits by making more holes within the boundaries of the array. See \cite{Fowler_2012} for a more detailed explanation.

\subsection{Distance of surface code}

In this subsection, we are going to demonstrate a way of finding the distance of a surface code in reference to finding the logical operators and then using that to create a set of operators that commutes with the stabilizer generator which in technical language is termed as the centralizers of stabilizers generator. As discussed in Section \ref{genstabcode}, the minimum weight of the elements of the centralizer of the stabilizer generator would provide the distance of the code. We defined the logical operator by creating a chain of operators commute with our stabilizer generator. As seen above in a simple 2D surface code there are 2 homologically inequivalent 1-cycles and hence two logical operators $\mathbf{X}_L$ and $\mathbf{Z}_L$. Fowler \emph{et al.} \cite{Fowler_2012} defined the distance of a 2D surface code as the minimum weight of the logical operators. For simple geometries in which we can encode only one qubit.

\section{Surface Code Implementation Using Cluster State}\label{result}

In the previous two sections, we discussed the concepts of surface codes and the measurement-based computation model utilizing cluster state generation separately. Now, in this section, we aim to explain whether we can study the stabilizer evolution and perform similar computations through a different initialization procedure that differs from the usual computational basis initialization of the qubit and CNOT entangling operation. To study a different initialization, we performed initialization of our qubits in the $|+\rangle$ which was followed by the CZ entangling operator in order to prepare the initial surface code cluster state. Such a circuit is used to study the stabilizer evolution of the prepared state and a comparison is demonstrated with respect to the usual computation basis surface code circuit in this section. Regarding the initialization, compared to the CNOT entangling operation, the CZ entanglement being symmetric allows one the flexibility of not worrying about defining the control and target qubits involved in the process while entangling the qubits to form the cluster state on which the computation can be performed. In the study to provide a comparison along with other computations, we will discuss our work wherein we realized a surface code on $3\times3$ and $5\times5$ cluster states. 

\subsection{Realizing Stabilizer Evolution for a Surface Code} 

To investigate a surface code that captures our attention, we employ Algorithm \ref{algo1}, a modified version of the algorithm proposed by \cite{shaw2023construction} for realizing stabilizer codes. This algorithm enables us to determine the ultimate stabilizer generator of the evolving cluster state. Subsequently, the cluster state undergoes a transformation using a suitable measurement pattern.  

Algorithm \ref{algo1} involves a meticulous selection process, wherein the stabilizer operator is carefully chosen to commute with every measurement operator during each 
\begin{algorithm}
\caption{An algorithm to realize stabilizer generators for a surface code}\label{algo1}
\begin{algorithmic}
\Require  Cluster State ($\mathcal{C}$), Measurement Array ($\mathcal{M}$), Non-Identity Measurement Index Array ($n_{m}$).

\State $\mathbf{K}_{in}^{(j)} = Stabilizer(\mathcal{C}) = \mathbf{X}_{a}\prod_{b \in N(a)} \mathbf{Z}_{b} $\Comment{a := elements of $\mathcal{C}$}

\For {$i$ in range(0,|$\mathcal{M}$|)}
    \State $\mathbf{K}_{c}$ , $\mathbf{K}_{ac}$ = $[$ $]$ 
        \For{$k$ in range(0,|$\mathbf{K}_{in}^{(j)}$|)}
            \If{$[ \mathbf{K}^{(j)}_{in}[k], \mathcal{M}[i] ] = 0$}
                \State $\mathbf{K}_{c} \gets \mathbf{K}_{in}^{(j)}[k]$
                \Else{}
                \State $\mathbf{K}_{ac} \gets \mathbf{K}_{in}^{(j)}[k]$
            \EndIf
        \EndFor

    \If{$|\mathbf{K}_{ac}|> 1$}
        \For{$l$ in range(0,|$K_{ac}$|-1)}
            \State $\mathbf{K}_{c} \gets \mathbf{K}_{ac}[l]\times \mathbf{K}_{ac}[l+1]$
        \EndFor
    \Else{}
        \State $\mathbf{K}_{ac}$ vanishes
    \EndIf

    \State $\mathbf{K}^{(j)} = [$ $]$
    \For{$m$ in range(0,|$\mathbf{K}_{c}$|)}
        \If{$[\mathbf{K}_{c}[m], \mathcal{M}[i]] = 0$}
            \If{$\mathbf{K}_c [m][n_{m}[i]]=\mathcal{M}[i][n_{m}[i]] \neq Pauli(I)$}
                \State $\mathbf{K}^{(j)} \gets \mathbf{K}_{c}[m]\times \mathcal{M}[i]$
                \Else{}
                \State $\mathbf{K}^{(j)} \gets \mathbf{K}_{c}[m]$
                \EndIf          
        \EndIf
        
    \EndFor
    
    \State $\mathbf{K}^{(j)}_{in}\gets \mathbf{K}^{(j)}$
\EndFor
\State \textbf{Post-measurement Stabilizer Generators:} $\mathbf{K}^{(j)}_{out} \gets \mathbf{K}^{(j)}_{in}$
\end{algorithmic}
\end{algorithm}
cycle specified by the measurement pattern. It depends on considering the cases of commutation relation between the evolving stabilizer generators (beginning from the initial cluster state generators) and the measurement pattern under consideration. At each cycle for an element of the measurement operator, we store the stabilizers in $\mathbf{K}_{c} $ or $\mathbf{K}_{ac}$ depending on whether the element commute or anti commute with measurement operator respectively. If the anticommuting array ($\mathbf{K}_{ac}$) contains more than one element, we multiply the elements within themselves and append the new terms in the commuting array (i.e., in $\mathbf{K}_{c}$) to get the final stabilizer generators corresponding to a particular measurement operator. We repeat the process for each element of the measurement array to get the final stabilizer for our cluster state. This ensures that the desired stabilizer generators  commute within themselves and with the prescribed measurement patterns.

The algorithm is implemented using the AWS Braket platforms and requires the Qiskit libraries. The resulting stabilizer generators obtained from this implementation will be utilized in the subsequent subsections, with an aim to demonstrate and study the observed changes compared to the established CNOT implementation for 3×3 and 5×5 circuits. 

\subsection{Surface Code Using 3\texorpdfstring{$\times$}{} 3 Cluster State}

In this section, we will conduct a comparable analysis on the stabilizer evolution of a 3×3 array of qubits, consisting of a cluster of 9 qubits, following a specific measurement pattern. This measurement pattern is an extension of the pattern utilized in the previous section. As illustrated in Figure \ref{3x3}, the measurement pattern involves \textbf{Z}-measurements on qubits ${1,7}$ and \textbf{X}-measurements on qubits ${3,5}$. The resulting circuit, which will be used for subsequent computations, takes the following form:
\begin{figure} [ht!]
\centering
\includegraphics[width=60mm]{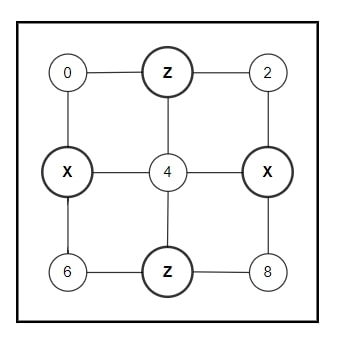}
\vspace{-0.5 cm}
\captionsetup{width=0.9\linewidth}
\caption{3$\times$3 cluster state. Qubits 1 and 7 are measured in $\textbf{Z}$-basis and qubits 3 and 5 are measured in $\textbf{X}$-basis.}
\label{3x3}
\end{figure}

\subsubsection{Stabilizer Evolution of a 3\texorpdfstring{$\times$}{}3 Surface Code:}On applying the algorithm to study the stabilizer evolution of the above-mentioned circuit with 5 data qubits and 4 measure qubits, as seen in Figure \ref{3x3}. We study the stabilizer evolution both for the CNOT entangled circuit and our CZ entangled circuit which can be seen in the following results:
\begin{figure}[ht!]
\centering
\hspace*{-1.8cm}
\includegraphics[width =190mm]{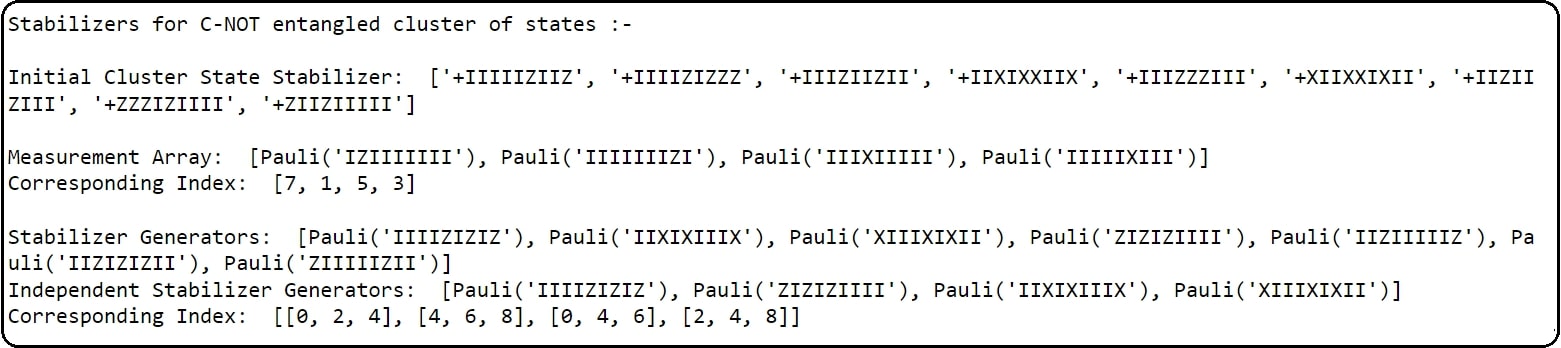}
\captionsetup{width=2.0\linewidth}
\caption{Stabilizer Generators for CNOT entangled 3x3 cluster.}
\label{ResultsCNOT}
\end{figure}
\vspace{-0.5 cm}
\begin{figure}[ht!]
\centering
\hspace*{-1.8cm}
\includegraphics[width =190mm]{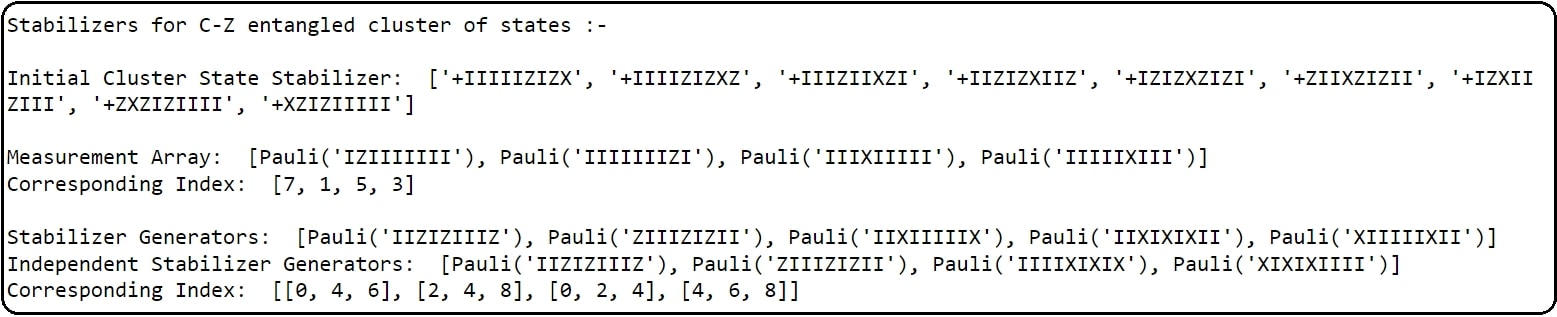}
\captionsetup{width=2.0\linewidth}
\caption{Stabilizer Generators for CZ entangled 3x3 cluster.}
\label{ResultsCZ}
\end{figure}
\\
\\
The above results demonstrate a deviation between the different surface code initialization mentioned using CNOT and CZ operations respectively. Opposite indexing for $\mathbf{Z}$ and $\mathbf{X}$ stabilizer is observed as a difference from the predecessor initialization which can be attributed to the fundamental Hadamard transformation incorporated in the theoretical framework as demonstrated in the next subsection demonstrating the theoretical calculation to support the results obtained. 

\subsubsection{Theoretical Background:}

From Figure \ref{3x3} we have $(3,5)\in \mathcal{C}_I(g)$, $(1,7)\in\mathcal{C}_{B}(g)$ and $(0,2,4,6,8)\in\mathcal{C}_O(g)$. Let the state of the cluster be $|\phi\rangle$. Using Equation \ref{stabilizers_cs}, we can write the following eigenvalue equations for $|\phi\rangle$:

\begin{align}
\mathbf{X}^{(0)}\mathbf{Z}^{(1)}\mathbf{Z}^{(3)}|\phi\rangle=|\phi\rangle,\\
\mathbf{X}^{(1)}\mathbf{Z}^{(0)}\mathbf{Z}^{(2)}\mathbf{Z}^{(4)}|\phi\rangle=|\phi\rangle,\\
\mathbf{X}^{(2)}\mathbf{Z}^{(1)}\mathbf{Z}^{(5)}|\phi\rangle=|\phi\rangle,\\
\mathbf{X}^{(3)}\mathbf{Z}^{(0)}\mathbf{Z}^{(4)}\mathbf{Z}^{(6)}|\phi\rangle=|\phi\rangle,\\
\mathbf{X}^{(4)}\mathbf{Z}^{(3)}\mathbf{Z}^{(1)}\mathbf{Z}^{(5)}\mathbf{Z}^{(7)}|\phi\rangle=|\phi\rangle,\\
\mathbf{X}^{(5)}\mathbf{Z}^{(2)}\mathbf{Z}^{(4)}\mathbf{Z}^{(8)}|\phi\rangle=|\phi\rangle,\\
\mathbf{X}^{(6)}\mathbf{Z}^{(3)}\mathbf{Z}^{(7)}|\phi\rangle=|\phi\rangle,\\
\mathbf{X}^{(7)}\mathbf{Z}^{(4)}\mathbf{Z}^{(6)}\mathbf{Z}^{(8)}|\phi\rangle=|\phi\rangle,\\
\mathbf{X}^{(8)}\mathbf{Z}^{(5)}\mathbf{Z}^{(7)}|\phi\rangle=|\phi\rangle.
\end{align}

\begin{figure}[ht!]
\centering
\includegraphics[width=50mm]{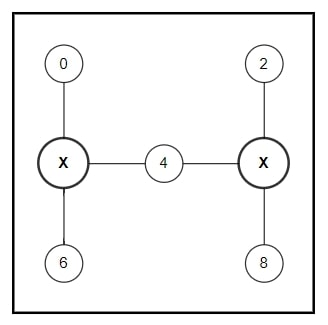}
\captionsetup{width=0.8\linewidth}
\vspace{-0.4 cm}
\caption{3$\times$3 cluster state after measurement of qubits 1 and 7 in $\mathbf{Z}$ basis. As seen in section \ref{effect_z} measuring in the $\textbf{Z}$-basis removes the qubits from the circuit.}
\label{after_z}
\end{figure}

Since qubits 1 and 7 are measured in $\mathbf{Z}$ basis we can remove these qubits from the state. This is because we saw in Section \ref{effect_z} that qubits measured in $\mathbf{Z}$ do not affect the final outcome and hence can be removed.  The remaining qubits are shown in Figure \ref{after_z}. Let the state of the remaining qubits be $|\psi\rangle$. It will satisfy the following set of cluster state eigenvalue equations:

\begin{align}
\mathbf{X}^{(0)}\mathbf{Z}^{(3)}|\psi\rangle=|\psi\rangle,\\
\mathbf{X}^{(2)}\mathbf{Z}^{(5)}|\psi\rangle=|\psi\rangle,\\
\mathbf{X}^{(3)}\mathbf{Z}^{(0)}\mathbf{Z}^{(4)}\mathbf{Z}^{(6)}|\phi\rangle=|\phi\rangle,\\
\mathbf{X}^{(4)}\mathbf{Z}^{(3)}\mathbf{Z}^{(5)}|\psi\rangle=|\psi\rangle,\\
\mathbf{X}^{(5)}\mathbf{Z}^{(2)}\mathbf{Z}^{(4)}\mathbf{Z}^{(8)}|\psi\rangle=|\psi\rangle,\\
\mathbf{X}^{(6)}\mathbf{Z}^{(3)}|\psi\rangle=|\psi\rangle,\\
\mathbf{X}^{(8)}\mathbf{Z}^{(5)}|\psi\rangle=|\psi\rangle.
\end{align}

Using the above equations we can write the following two equations:
\begin{align}
\mathbf{Z}^{(3)}\mathbf{Z}^{(5)}\mathbf{H}^{(0)}\mathbf{H}^{(4)}\mathbf{H}^{(6)}\mathbf{Z}^{(0)}\mathbf{Z}^{(4)}\mathbf{Z}^{(6)}\mathbf{H}^{\dagger(0)}\mathbf{H}^{\dagger(4)}\mathbf{H}^{\dagger(6)}|\psi\rangle=|\psi\rangle,\\
\mathbf{Z}^{(3)}\mathbf{Z}^{(5)}\mathbf{H}^{(2)}\mathbf{H}^{(4)}\mathbf{H}^{(8)}\mathbf{Z}^{(2)}\mathbf{Z}^{(4)}\mathbf{Z}^{(8)}\mathbf{H}^{\dagger(2)}\mathbf{H}^{\dagger(4)}\mathbf{H}^{\dagger(8)}|\psi\rangle=|\psi\rangle.
\end{align}
Now, by incorporating Theorem 1, we can say that we have to apply an additional Hadamard gate ($\mathbf{H}$) on the output qubits (0,2,4,6,8) to get the correct surface code stabilizers. 
\subsection{Surface Code Using 5\texorpdfstring{$\times$}{}5 Cluster State}
In this section, we will perform a similar study of the stabilizer evolution of a cluster of 25 qubits stacked as a 5$\times$5 array of qubits following a particular measurement pattern. As seen in Figure \ref{5x5},  the measurement pattern followed is an extension of the pattern with \textbf{Z}-measurement and \textbf{X}-measurement being performed on qubits $\{1,3,11,13,21,23\}$ and $\{5,7,9,15,17,19\}$ respectively, with the circuit to be considered taking the following form:
\vspace{-0.4 cm}
\begin{figure}[ht!]
\centering
\includegraphics[width=70mm]{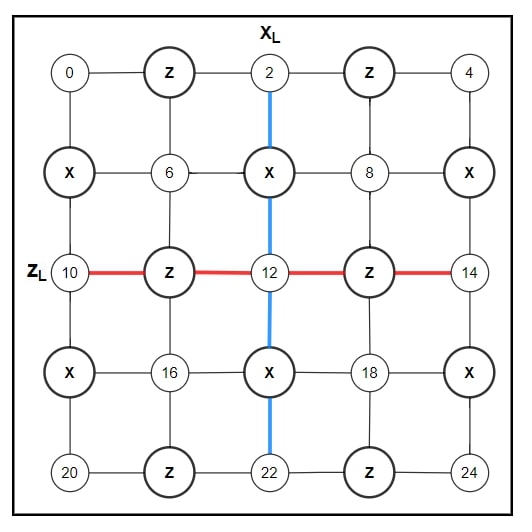}
\vspace{-0.2 cm}
\captionsetup{width=1.0\linewidth}
\caption{5$\times$5 cluster state showing the measurement pattern implemented to obtain the surface code. The logical operators $\hat{\textbf{X}}_L$ (red) and $\hat{\textbf{Z}}_L$ (blue) are also shown.}
\label{5x5}
\end{figure}
\subsubsection{Stabilizer Evolution of a 5\texorpdfstring{$\times$}{}5 Surface Code:}
On applying the algorithm to study the stabilizer evolution of the above-mentioned circuit with 13 data qubits and 12 measure qubits, we obtain the following stabilizer evolution for our circuit:
\begin{figure}[ht!]
\centering
\hspace*{-1.8cm}
\includegraphics[width =190mm]{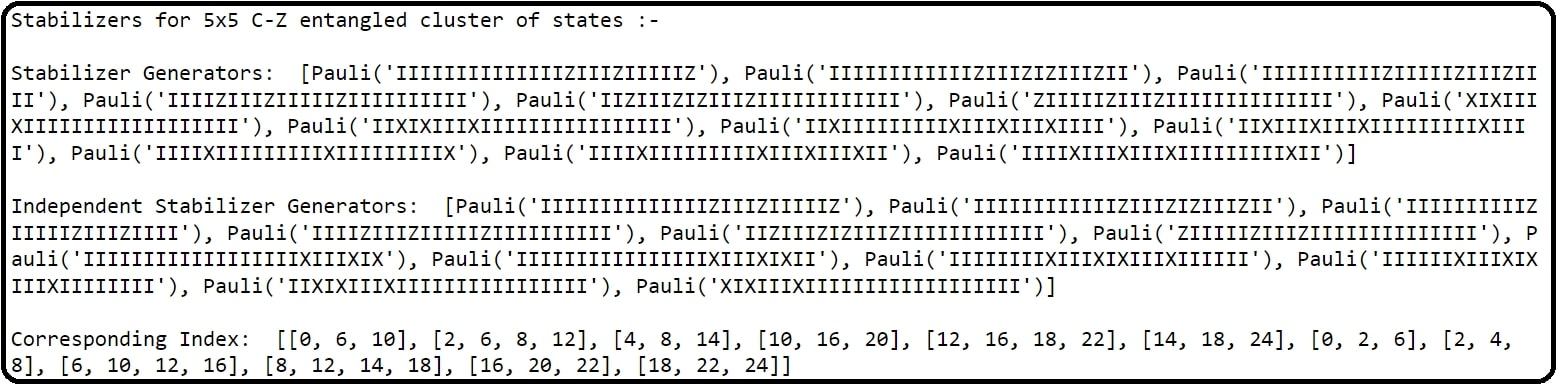}
\captionsetup{width=2.0\linewidth}
\caption{Stabilizer Generators for CZ entangled 5x5 cluster.}
\label{Results5x5CZ}
\end{figure}

On studying the corresponding indexes in Figure \ref{Results5x5CZ} of our independent stabilizer generator, a similar stabilizer evolution of different indexing is observed in this case as well. The opposite indexing of the stabilizer as compared to its equivalent CNOT implementation counterpart can be similarly explained by extending the theoretical background mentioned in the study of a 3$\times$3 surface code.  

\subsubsection{Distance of a 5\texorpdfstring{$\times$}5 Surface Code:}

In this subsection, we are going to demonstrate a way of finding the distance of a surface code in reference to finding the logical operators and then using that to create a set of operators that commutes with the stabilizer generator which in technical language is termed as the centralizers of stabilizers generator. 
As discussed in Section \ref{genstabcode}, the minimum weight of the elements of the centralizer of the stabilizer generator would provide the distance of the code. We defined the logical operator by creating a chain of operators commute with our stabilizer generator. In the process, we make an observation that the chain of Pauli Operators that constitutes the logical operators also follow similar opposite indexing of \textbf{Z}-logical and \textbf{X}-logical operators in a similar fashion as the stabilizer generators of our surface codes which can be attributed once again to the fundamental Hadamard transformation involved in the theoretical background of our cluster state formulation. Taking care of the opposite indexing, creating the logical operators (the first three elements in the centralizer array below corresponds to the \textbf{X}, \textbf{Z}, and \textbf{Y} logical operator respectively), and performing multiplications of logical operators and stabilizer generators gives us our required element of the centralizers of the stabilizer along with the corresponding weight which is helpful to find the distance of the code, as demonstrated below:

\begin{figure}[ht!]
\centering
\hspace*{-1.8cm}
\includegraphics[width =190mm]{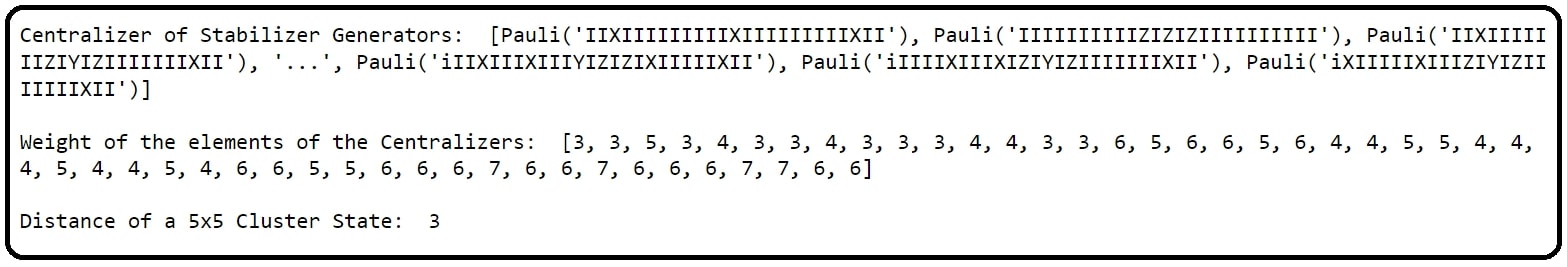}
\captionsetup{width=2.0 \linewidth}
\caption{Centralizers for a 5$\times$5 Surface Code.}
\label{cent}
\end{figure}

As seen from Figure \ref{cent}, the distance of the code is 3, which implies that our 5x5 surface code can be denoted as $[[13,1,3]]$. With such a code of distance 3, we can hope to correct 1 error. Such a structure provides the smallest structure which can be used as an error-correcting code for future consideration of the study. The work demonstrates how one using such an initialization of a surface code structure can result in just the opposite indexing of the stabilizer generator which can be taken care of by using a suitable Hadamard transformation while providing easier initialization to our CNOT entangled cluster state.

\section{Discussion}

In this paper, we proposed a way to create surface codes using cluster state as the resource. We noted that the cluster state, formed using the CZ operation, gives an advantage over the entangled state formed using the CNOT operation. This is primarily because the CZ operation is symmetric and does not differentiate between the control and target qubits. Another benefit of using cluster states was that it is practically easy to create in the lab. The unitary interaction, in Equation \ref{unitary_int}, corresponds to an Ising-type interaction that can be realized via controlled collisions in optical lattices \cite{raussendorf2001one}. This means that the temporal resources required to create a cluster state are independent of the size of the cluster. Once the cluster states are obtained we applied MBQC to get the desired surface code. 

We presented an algorithm to obtain the stabilizers of the surface codes. Stabilizer formalism plays a central role in all of quantum information specifically in fault-tolerant quantum computation. We only analyzed simple 2D square-shaped surface code structures mainly 3$\times$3 and 5$\times$5. With the help of stabilizers, we were able to obtain the distance. The geometry of our surface codes only allowed two homologically inequivalent logical operators; hence, we could only encode one logical qubit irrespective of the number of physical qubits.  We noted that the distance of the surface code can be written as a function of $n$ as $d=\frac{n+1}{2}$. We further saw that the 3$\times$3 surface is of no practical use as it cannot correct any quantum error. But 5$\times$5 can be used to correct one type of error in the encoded logical qubit. From our analysis, we can see that MBQC provides an efficient mechanism to create surface codes. These surface codes can then be used as an encoder circuit to encode one or more than one logical qubit depending on the topology and structure of the surface code. 

\section{Acknowledgement}

We thank MeitY QCAL for their generous financial support, which made this research possible. Dr. A.R. thanks the IISER-B for the startup fund for the same. We would also like to acknowledge Mr. Mehul Chakraborty's contributions to the discussions during the project's initial phase. 
\pagebreak
\section{Appendix}

\subsection{\textbf{3 qubit code }}\label{3qucode}

Now, let's see how we can devise a 3-qubit stabilizer code using the method discussed in Section \ref{qecc}.
We know that in 3 qubit code three physical qubits are used to encode one logical qubit, hence $n=3$ and $k=1$. Qubit flip errors in one qubit are described by the following error operators
\begin{equation}\label{error_op}
\mathcal{E}=\left\{\mathbf{X}\otimes\mathbf{I}\otimes\mathbf{I}, \mathbf{I}\otimes\mathbf{X}\otimes\mathbf{I}, \mathbf{I}\otimes\mathbf{I}\otimes\mathbf{X}\right\}.
\end{equation}
We select from all the three-qubit Pauli operators elements that commute with themselves and anti-commute with elements of $\mathcal{E}$ in Equation \ref{error_op}. We will obtain the following set of stabilizers
\begin{equation}
\mathbb{S}=\left\{\mathbf{Z}\otimes\mathbf{Z}\otimes\mathbf{I}, \mathbf{I}\otimes\mathbf{Z}\otimes\mathbf{Z}, \mathbf{Z}\otimes\mathbf{I}\otimes\mathbf{Z}\right\}.
\end{equation}
Any 2 elements of $\mathbb{S}$ form the generator of the set. Therefore the code subspace has dimensions 2 and has the eigenstates
\begin{equation}
\begin{aligned}
|0\rangle_L=|000\rangle,\\
|1\rangle_L=|111\rangle.
\end{aligned}
\end{equation}
\subsection{\textbf{5 qubit code}}\label{5qucode}

Table \ref{stabilizer_fiveq} shows the stabilizer group generators for a five-qubit code. Since no. of generators is 4, i.e., $r=4$, we have $k=n-r=1$. Consider an error $\mathbf{E}=\mathbf{Y}\otimes\mathbf{Z}\otimes\mathbf{Y}\otimes\mathbf{I}\otimes\mathbf{I}$. This error commutes with all the generators. Therefore, $\mathbf{E}\in\mathbb{N}_\mathbb{S}\backslash \mathbb{S}$. We will see that this is the smallest error that is undetectable; hence this code has distance $d=3$. One can also note that this is a \emph{perfect} code; it means that all the possible single-qubit errors (16 in this case) exhaust all the possible error syndromes (also 16). In Table \ref{stabilizer_fiveq}, $\textbf{X}$ and $\textbf{Z}$ define the encoded bit-flip and phase-flip operators respectively. Now, using the stabilizer group we can define the basis for the code subspace
\begin{equation}
\begin{aligned}
|\bar{0}\rangle&=\sum_{\textbf{M}\in\mathbb{S}}\textbf{M}|00000\rangle,\\
|\bar{1}\rangle&=\textbf{X}|\bar{0}\rangle.
\end{aligned}
\end{equation}
\begin{table}[ht!]
\centering
\begin{tabular}{c|ccccc}
$\textbf{M}_{1}$ & $\textbf{X}$ & $\textbf{Z}$ & $\textbf{Z}$ & $\textbf{Z}$ & $\textbf{I}$ \\
$\textbf{M}_{2}$ & $\textbf{I}$ & $\textbf{X}$ & $\textbf{Z}$ & $\textbf{Z}$ & $\textbf{X}$ \\
$\textbf{M}_{3}$ & $\textbf{X}$ & $\textbf{I}$ & $\textbf{X}$ & $\textbf{Z}$ & $\textbf{Z}$ \\
$\textbf{M}_{4}$ & $\textbf{Z}$ & $\textbf{X}$ & $\textbf{I}$ & $\textbf{X}$ & $\textbf{Z}$ \\
\hline 
$\textbf{X}_L$ & $\textbf{X}$ & $\textbf{X}$ & $\textbf{X}$ & $\textbf{X}$ & $\textbf{X}$ \\
$\textbf{Z}_L$ & $\textbf{Z}$ & $\textbf{Z}$ & $\textbf{Z}$ & $\textbf{Z}$ & $\textbf{Z}$
\end{tabular}
\caption{Stabilizer for five-qubit code $[[5,1,3]]$.}
\label{stabilizer_fiveq}
\end{table}

\subsection{\textbf{Homology of curves}}\label{homology}
Homology is a branch of topology that is most relevant for topological codes. Topological codes can be formulated and understood almost entirely in homological terms. Homology is the mathematical field that abstracts and generalizes the notion
of boundary. In this work, we will limit our study to homology on the group $\mathbb{Z}_2$. We will see that there is a topological invariant called the $\emph{first homology group}$ of the surface which is essential in making topological codes \cite{dan_b}.

\subsubsection{\texorpdfstring{$\mathbb{Z}_2$}{} homology}
\begin{figure}[ht!]
\centering
\includegraphics[width=85mm]{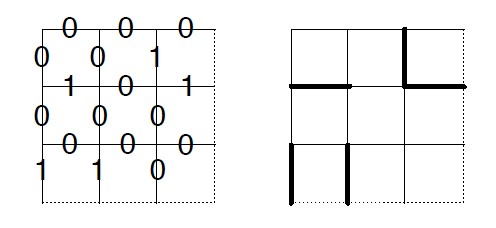}
\captionsetup{width=1.0\linewidth}
\caption{1-chain in $\mathbb{Z}_2$ homology \cite{dan_b}. (\emph{Permission Granted by the author})}
\label{1_chain}
\end{figure}

\begin{figure}[ht!]
\centering
\includegraphics[width=110mm]{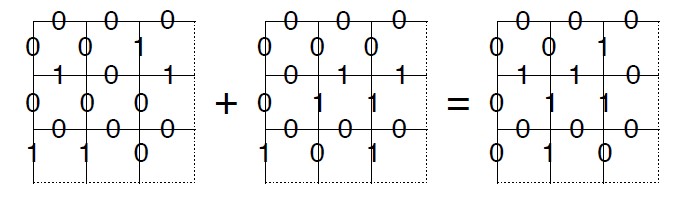}
\captionsetup{width=0.8\linewidth}
\caption{Addition of 1-chains in $\mathbb{Z}_2$ homology \cite{dan_b}. (\emph{Permission granted by the author})}
\label{1_chain_add}
\end{figure}

Consider a particular lattice embedded in the surface. We write vertices as 0-cells, edges as 1-cells, and faces as 2-cells. We will associate an element from the $\mathbb{Z}_2$ group with each $n$-cell, with $n$=0,1,2 for our case. The group $\mathbb{Z}_2$ is the simplest non-trivial group. It has two elements and is isomorphic to the set of integers $\{0,1\}$ with group composition given by addition modulo 2. For example, if we label edges as $\{e_i\}_1^E$ we can represent any set of edges $E^{\prime}$ as
\begin{equation}
c=\sum_ic_ie_i\qquad c_i=\begin{cases}
0 & \text{if}\;e_i\notin E^{\prime}\\
1 & \text{if}\;e_i\in E^{\prime}
\end{cases}.
\end{equation}
A sum of this form is called a 1-chain. Given any set of edges $E^{\prime}$ we can represent 1-chain by substituting each element in $E^{\prime}$ by 1 and the rest of the elements by 0. One such example is shown in Figure \ref{1_chain}.
\\
We can add two 1-chains via bit-wise modulo 2 addition to produce another 1-chain (See Figure \ref{1_chain_add}). We can easily see that 1-chain forms an abelian group under component-wise addition modulo 2. We denote this group by $ \mathbb{C}_1\simeq \mathbb{Z}_2^E$ where $E$ is the number of edges. This means that $\mathbb{C}_1$ is the direct product of $\mathbb{Z}_2$ done $E$ times. The order of the group is $2^E$. Further, we identify $E$ independent generators of the group by associating a generator with each edge. We place 1 on the edge associated with a specific generator and 0 on every other edge. This procedure gives us $E$ generators, also called edge generators. Similarly, we can form the group of 0-chains $\mathbb{C}_0\simeq\mathbb{Z}_2^V$ and the group of 2-chains $\mathbb{C}_2\simeq\mathbb{Z}_2^F$. One can also visualize $\mathbb{C}_i$ as vector spaces with group generators being one set of basis elements $B(\mathbb{C}_i)$.

\subsubsection{Boundary operators}
We introduce a family of group homomorphisms $\partial$ called boundary operators or maps. As the name suggests, $\partial$ takes objects to their boundaries. We define a homomorphism $\partial_i:\mathbb{C}_{i}\rightarrow \mathbb{C}_{i-1}$ such that:
\begin{equation}
\partial_{i-1}\cdot\partial_i=0.
\end{equation}
$\partial_2$ (2-boundary operator/map) maps a set of faces to a set of edges that forms its boundary. $\partial_1$ (1-boundary operator/map) maps a set of edges to a set of vertices where an odd number of edges meet. We can define 0-boundary operator $\partial_0$ as a map that takes every 0-chain to null chain, i.e., $\partial_{0}c=0$. Null chain is sometimes referred to as 1-chain.

\subsubsection{Cycles}
We define a subgroup $\mathbb{Z}_1\subset \mathbb{C}_1$ which is a group of 1-chains $z$ that have no boundary, that is $\partial_1z=0$, the kernel of $\partial_1$. Elements of $\mathbb{Z}_1$ are also called 1-cycle. Generally, an $n$-cycle is an $n$-chain with a null boundary. The group of $n$-cycles is labeled as $\mathbb{Z}_n$. Now, a crucial observation is that all boundaries are also cycles. We define a subgroup $\mathbb{B}_1\subset \mathbb{C}_1$ which is a group of 1-chains $b$ that are a boundary of a 2-chain $c$, i.e., $\partial_2c=b$ or $b\in\operatorname{img}(\partial_2)$ for some $c\in \mathbb{C}_2$. This means that 
\begin{equation}
\partial^2c:=\partial_1\cdot\partial_2c=0\quad\forall c\in \mathbb{C}_2.
\end{equation}
\subsection{Homological Equivalence}\label{homological_equivalence}
\begin{figure}[ht!]
\centering
\includegraphics[width=100mm]{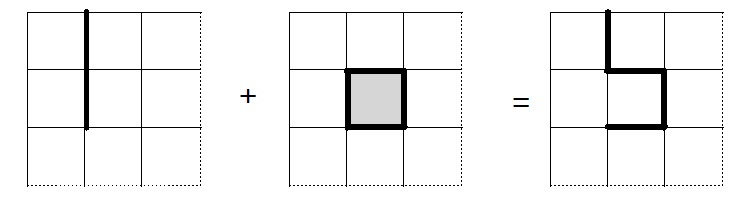}
\captionsetup{width=0.8\linewidth}
\caption{Homologically equivalent 1-chains \cite{dan_b}. (\emph{Permission granted by the author})}
\label{homo_equi}
\end{figure}

Two $n$-chains are homologically equivalent if they are equal up to composition with an $n$-boundary. This means if two chains $c$ and $c^{\prime}$ are homologically equivalent, then there exists a ($i+1$)-chain $c_{i+1}$ such that
\begin{equation}
c_i=c_i^{\prime}+\partial c_{i+1}.
\end{equation}
Fig \ref{homo_equi} shows two homologically equivalent 1-chains.
\subsection{Homology group}
If two chains are homologically equivalent then they share the same boundary but the reverse is not always true, that is, not all chains with the same boundary are homologically equivalent. Homological equivalence partitions the subgroup of chains with a given
boundary into equivalence classes. Since cycles are the chains with null boundary it can be partitioned under homological equivalence. We define a quotient group $\mathbb{H}_n$ by partitioning $\mathbb{Z}_n$ into equivalence classes under composition with elements of $\mathbb{B}_n$ as
\begin{equation}
\mathbb{H}_n=\mathbb{Z}_n/\mathbb{B}_n.
\end{equation}
From algebraic topology, we know that the properties of the homology group depend only on the topology of the surface it is defined. The first homology group, up to isomorphisms, can be written as
\begin{equation}\label{first_homology_group}
\mathbb{H}_1=\mathbb{Z}_1/\mathbb{B}_1\simeq\mathbb{Z}_2^{2g}.
\end{equation}
The elements of $\mathbb{H}_1$ are cosets of the form $\bar{z}:=\{z+b|b\in B_1\}$ for some $z\in \mathbb{Z}_1$.

\end{document}